# Totally secure classical networks with multipoint telecloning (teleportation) of classical bits through loops with Johnson-like noise[1]

LASZLO B. KISH[+] AND ROBERT MINGESZ[#]

[+]*Department of Electrical Engineering, Texas A&M University, College Station, TX 77843-3128, USA*

[#]*Department of Experimental Physics, University of Szeged, Dom ter 9, Szeged, H-6720, Hungary*

(Versions: March 5, 7, 8, 9, 13, 17, 20, 21, 2006)

First, we show a new inexpensive defense against intruders and the man-in-the-middle attack in the Kirchhoff's-loop-Johnson-like-noise (KLJN) cipher. Then instead of point-to-point communication, we propose a high efficiency, secure network. The (in the idealistic case totally secure) classical network is based on an improved version of the KLJN cipher. The network consists of two parallel networks: *i)* a chain-like network of securely communicating, electrically isolated Kirchhoff-loops with Johnson-like noise and driven by a specific switching process of the resistances; ii) and a regular non-secure data network with a Coordinator-server. If the classical network is fast enough, the chain-like network of N communicators can generate and share an N bit long secret key within a *single clock period* of the ciphers and that implies a significant speed-up compared to the point-to-point key exchanges used by quantum communication or RSA-like key exchange methods. This is a teleportation-type multiple telecloning of the classical information bit because the information transfer can take place without the actual presence of the information bit at the intermediate points of the network. At the point of the telecloning, the clone is created by the product of a bit coming from the regular network and a secure bit from the local KLJN ciphers. The same idea could be implemented with quantum communicator pairs producing entangled secure bit at the two ends. This is the telecloning of classical bits via quantum communicator networks without telecloning the quantum states.

## 1. Introduction: totally secure statistical physical communicator with Kirchoff-loop and Johnson(-like) noise

Recently, a totally secure classical communication scheme was introduced [1,2] utilizing two identical pairs of resistors and noise voltage generators, the physical properties of an idealized Kirchoff-loop and the statistical physical properties thermal noise, see Figures 1 and 2. The resistors (low bit = small resistor $R_S$, high bit = large resistor $R_L$) and their noise voltage generators are randomly chosen and connected at each clock period at the two sides of the information channel. A *secure bit exchange* takes place when the states at the two ends are different, which is indicated by an intermediate level of the *rms* noise voltage on the line, or that of the *rms* current noise in the wire. The most attractive properties of the KLJN cipher are related to its *security* [1, 3-5] and the *robustness* of classical information. In the *idealized* scheme of the Kirchoff-loop-Johnson(-like)-noise (KLJN) cipher, the passively observing eavesdropper can extract *zero bit* of information. The total security of the idealized system is based on Statistical Physics, and inherently connected to the Second Law of Thermodynamics

---

[1] This paper has been pre-published and updated at arxiv.org since March 5, 2006; http://arxiv.org/abs/physics/0603041.

(Fluctuation-Dissipation Theorem), the Energy Conservation Law (Kirchhoff's Loop-Law) and the rules of algebraic linear equation systems (N independent equations are needed to determine N unknown variables).

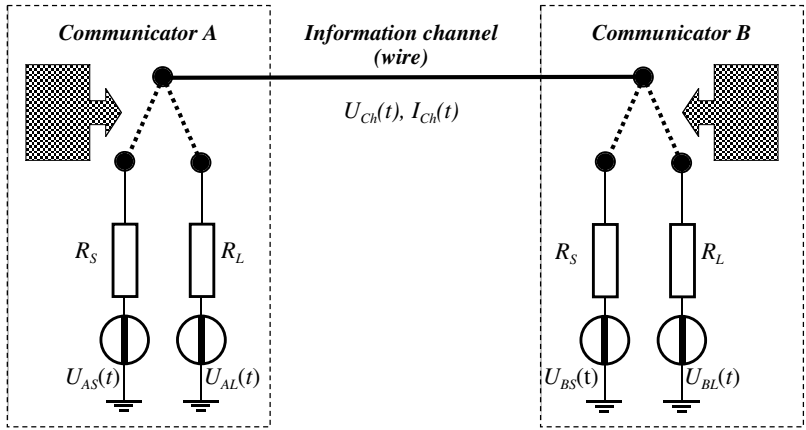

**Figure 1.** Secure key generation and distribution via the Kirchhoff-loop-Johnson(-like)-noise (KLJN) cipher (idealized picture). The joint information of the two end-bits is carried by the statistics of the *rms* noise voltage and current in the channel. Secure bit exchange takes place whenever the end-bits are different (intermediate value of the *rms* noise voltage and current). To extract the secure information bit of one end, the knowledge of this joint information and that of the information bit at the other end is needed. The passive eavesdropper can access only the joint information, so she knows that the bits are opposite, but she has zero information about the actual end bit.

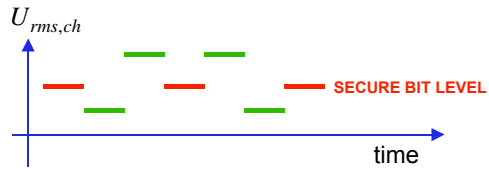

**Figure 2.** Channel noise *rms* voltage level due to the random bits at the two ends of the line; and secure bit exchange in the KLJN cipher. The *rms* value is obtained through a single clock period. On the average, only 50% of the clock periods can exchange secure bits. The eavesdropper does not know which end of the line has the low and which one has the high bit.

## 2. On the security of the KLJN cipher

Though there have been many comments about the fact that the practical cipher is not totally secure (see their refusals [4,5] and a correct point [6]), the *total security of the idealized (mathematical model) system remains unchallenged*. The ideality of the KLJN cipher means that the physical/circuit model [1] shown in Figure 1 *must be* the correct description of the situation. This requirement implies an ultimate high-frequency limit [1] for any normal or parasite frequency components in the channel to avoid wave affects (delay, propagation, reflection, etc.). This "no-wave" bandwidth limit is scaling with the reciprocal of the wire length [1]. Moreover, any other deviations from the model in Figure 1, such as parasite elements or other modification pose the risk of information leak [3-6]. The most significant, practical, non-ideality problems are the cable resistance pointed out by many commenters, first by Janos Bergou [6] and the cable capacitance [3]. However, by using sufficiently thick wires, these problems can be eliminated and the idealized situation and total security can *arbitrarily* be *approached* depending on the *available resources*. The situation is similar to the case of quantum communication where the theoretical security can only be approached but can never be reached due to the lack of ideal single-photon source, and the noise in the cannel and the detectors.

## 3. The earlier and new, economical solutions for defense against intruders and the man-in-the-middle attack

The *intrusive eavesdropper*, who emits a large and short current pulse in the channel, can extract *only one bit* of information while she is getting discovered [1]. The KLJN cipher with public channel for comparing currents is naturally protected against the *man-in-the-middle* (MITM) attack [3] and the eavesdropper is discovered with a very high probability while or before she can extract a single bit. Enhanced security can be reached by comparing the voltages and then the eavesdropper is discovered with a very high probability before she can extract a single bit of information [3]. The fastest key distribution with the highest security needs these defenses, see Sections 7 and 8. However as an economical solution, it is possible to detect the eavesdropping using only the statistical analysis of the noise (similarly to quantum communication) however that method would allow the eavesdropper to extract a number of bits before she gets discovered, *just like at quantum communication*. On the other hand, in the case of the man-in-the middle attack, a 4-bits average security can be reached if the parties communicate via the public channel only the fact of secure bit exchange, which is a satisfactory and inexpensive solution. It is so because the man-at-the-middle cannot predict the time when the end bits and its own bits will favor a secure bit pair alignment. Thus, on the average, after 4 bits one communicator will see secure transfer but the other one does not, consequently their exchanged public information will contradict and the "eavesdropping alarm" goes on. This economical solution requires exchanging the same number of bits via the secure and the public channel. However, see Sections 7 and 8, the unconditional security will require extra clock steps which will slow down the key generation and distribution by a factor of 2 or 3.

## 4. The enhanced KLJM communicator with complementary loops

The original KLJN cipher exchanges secure bits only 50% of the time. This property can cause an exponential slow-down versus distance if the information is measured and further transmitter in a network of coupled KLJN ciphers. In the rest of this paper, first we show an enhanced KLJN communicator which exchanges bits at 100% of the time. Then we show a network solution where the intermediate network elements do not have to decode the information in order to forward it. Because the information "gets through" a chain of transmitters without even being there, the mechanism is a *classical teleportation* [7] effect, more precisely, *telecloning* because the information will be present at both the source and the receiver Units. In this paper, we will apply the following convenient mathematical fashion. We will represent the "high" information bit by +1 and the "low" information bit by -1. Thus if $A$ is the logic state of a port driving one end of a KLJN cipher then the inverse satisfies $\bar{A} = -A$.

The enhanced KLJN communicator makes use of each clock period by using two KLJN ciphers driven in anticorrelated way, see Figure 3. This lossless KLJN communicator consists of two *electrically isolated* parallel KLJN loops with ports driven in the same fashion at one end and in a complementary way at the other end. Then one of the loops will always communicate a secure bit because the logical state will be opposite at the ends of one of the loops.

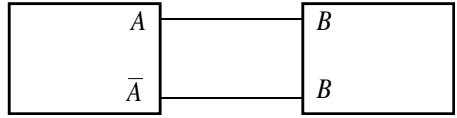

**Figure 3.** Lossless unsecure KLJN communication with two parallel ports. Each clock period transfers a bit. However the data are not secure as soon as the eavesdropper learns the port arrangement.

An important weakness is that the communicator shown in Figure 3 is not secure if the parties must agree through a regular network about who is running the correlated and who does the complementary ports. As soon as the eavesdropper learns the port arrangement, she can apparently decode the bit status at the two ends.

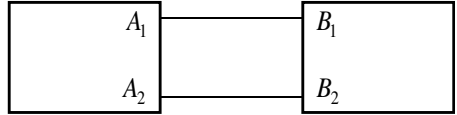

**Figure 4.** The Secure Complementary (SC) cipher and the secure distribution of the port control.

However, by using the totally secure nature of the KLJN communication, it is possible to arrange the port distribution in an unconditionally secure way; let us call this Secure Complementary (SC) cipher. The secure sharing of the parallel KLJN ports of Communicators (*A*) and (*B*) is done one time only, at the initialization. The ports $A_1$ and $B_1$ are randomly driven (connecting one of their resistors) at each clock period until a secure bit is exchanged between them (intermediate *rms* noise voltage level). Then, for example, they can use the following pre-agreed table for sharing:

| Situation of secure bits | Low (-1) at $B_1$ High (1) at $A_1$ | Low (-1) at $A_1$ High (1) at $B_1$ |
|---|---|---|
| Resulting port control | $A_1 = A_2$ $B_1 = \bar{B}_2 \equiv -B_2$ | $A_1 = \bar{A}_2 \equiv -A_2$ $B_1 = B_2$ |

**Table 1.** Truth table of sharing the port controls at the SC cipher.

The SC communicator is also lossless and it communicates a secure bit at each clock period with total security in the idealized case, see Figure 5.

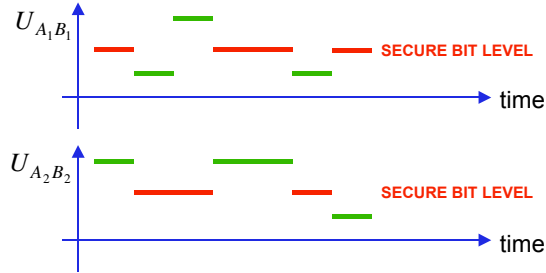

**Figure 5.** Example of channel noise *rms* voltage levels at lossless KLJN communication. All clock periods provide a secure exchange of (inverse) bits via one of the lines. Because the eavesdropper does not know the port control rules, she cannot access the information.

## 5. Networks: initialization and high-speed bit-exchange via telecloning

In this section, we describe the network and a simple initialization process. However, this initialization process allows certain vulnerabilities against the "Mingesz attacks" (see Section 7) especially during the first network-key generation run. In Section 7, we shall describe the more time-demanding but totally secure initialization process. In Section 8, we describe the most economical and still satisfactory solution.

The KLJN network consists of electrically isolated Kirchhoff loops driven in a specific way and a regular network with a special Coordinator-server. For the sake of simplicity, but without the restriction of generality, we consider a one-dimensional (chain-like) network where the network Units are connected only to their two nearest neighbors, see Figure 6.

First, we describe the KLJN cipher Units of the network. The connection between the network Units is made by SC cipher channels, see Figure 7. One SC connection is to the left hand neighbor via the left hand side ports $L$ and the other one is to the right hand side neighbor via the ports $R$. The initialization of the units is described below.

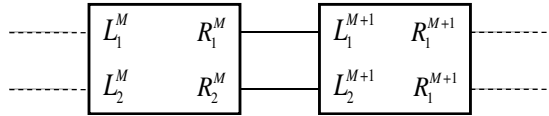

**Figure 6.** The *M*-th and *(M+1)*-th Units of the one-dimensional network and their external connections.

*i)* In the simplest case the initialization is similar to the *SC* cipher initialization described above (see also Sections 7,8). This is done one time only if the units are defended by the current/voltage method [3] (see Sections 7,8). The ports $R_1^M$ and $L_1^{M+1}$ of Units (*M*) and (*M*+1) are randomly driven (connecting one of their resistors) at each clock period until a secure bit is exchanged between them (intermediate *rms* noise voltage level). Then, for example, they can use the following pre-agreed table for sharing:

| Situation of secure bits | Low (small resistor) at $L_1^{M+1}$ High (large resistor) at $R_1^M$ | Low (small resistor) at $R_1^M$ High (large resistor) at $L_1^{M+1}$ |
|---|---|---|
| Resulting port control | $R_1^M = R_2^M$ $L_1^{M+1} = -L_2^{M+1}$ | $R_1^M = -R_2^M$ $L_1^{M+1} = L_2^{M+1}$ |

**Table 2.** Truth table of sharing the port controls between two nearest neighbors in the chain-like KJLN network.

Now, let us see how the network works. The Units are connected via the secure lossless KLJN chain described above and they communicate through the regular network, too, see Figure 7. After the initialization described above, the Units generate random bits so that they satisfy their port control initialization rules and the actual internal logic connection rule (regenerated at each clock period). The Units report the

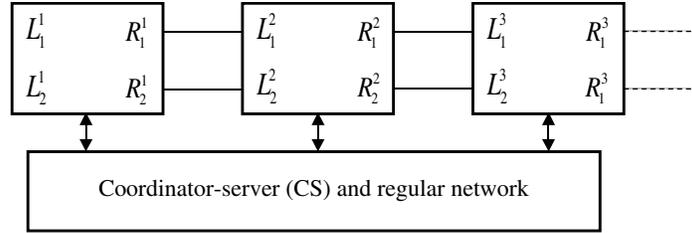

**Figure 7.** KLJN cipher based network for high-speed secret key generation and distribution via telecloning, with 100% fidelity, through chains of electrically isolated Kirchhoff loops. This system provides the highest speed when the defense is current/voltage based [3], see Sections 7 and 8. Note, a very good alternative can be realized with single wire R-L connections and regular KLJN ciphers (see Section 8).

logic relation $F$ (+1 or -1) between their secure bit at their $L$-ports and their secure bit at their R ports to the Coordinator-server (CS) of the regular network. Let us suppose that Unit (N) wants to clone the actual secure bit at the left hand side of Unit (M), where M<N. Then Unit (N) sends this request to the CS. The CS calculates

$$G_{M,N} = (-1)^{N-M} \prod_{j=M}^{N-1} F_j , \qquad (1)$$

where $G_{M,N}$ is the logic relation between the secure bit at the $L^M$ ports of Unit (M) and the secure bit at the $L^{(N)}$ ports of Unit (N); and $F_i$ is the reported logic relation between the secure bit at the $L^j$ ports and secure bit at the $R^j$ ports of Unit (*j*). Thus Unit (N) will make the following calculation:

$$X_L^M = G_{M,N} X_L^N , \qquad (2)$$

where $X_L^M$ is the secure bit at the $L^M$ ports of Unit (M) and $X_L^N$ is the secure bit of the $L^N$

ports of Unit (N). Because Unit (N) knows the secure bit values at his own ports and because he receives the $G_{M,N}$ information from the CS, he can evaluate Equation 3 and that is equivalent with the (tele)cloning of the $X_L^M$ secure bit with 100% fidelity. Let us see, why this process is a telecloning. The $X_L^M$ clone is the product of two information bits, the $G_{M,N}$ arriving via the regular network from CS and the local secure bit $X_L^N$. Therefore, the information does not exist on the chain of intermediate Units. It looks like it is "teleported" via the chain built of electrically isolated Kirchoff loops. However, because the $X_L^M$ bit stays intact at its original location during this operation, the proper term is telecloning, not teleportation.

We would like to mention that the expressions of "teleportation" and "telecloning" of information arise from the research on quantum communication (see for example [8,9]) and they had been considered as the unique properties of quantum systems, just like secure physical layers for communications. However, recently it was shown by Oliver Cohn [10] that classical teleportation of classical information is also possible. In his model, he was using boxes and coins and reached the teleportation of information by a total, 100%, fidelity. In the present paper we have reported telecloning of information also with total, 100%, fidelity, which is impossible to reach in a quantum system due to fundamental mathematical constraints [9] of the telecloning operations and due to the extreme fragility of quantum information.

## 6. High-speed secret key distribution over the whole network

Because the CS collects all the logic relations between the secure bits at the *L* and *R* ports of each Units, CS can provide the relevant $G_{k,m}$ bit for Unit (*m*) to teleclone the secure bit at the *L* ports of Unit (*k*). This information can be given to all Units simultaneously about all the other Units. In this way, each Unit can teleclone all the secure bits at the other Units, for each clock period, provided the regular network is fast enough. That means, if the network consist of *N* Units, an (*N*-2) bit long secret key is generated and securely distributed over the whole network *at each clock period*. (We supposed that the L ports of Unit (1) and the R ports of the last Unit in the chain are not connected; this is the reason for the -2 correction of N). If the regular network is slow, the *X* bits can be recorded by the Units and they can make the telecloning later when they receive the relevant *G* bits.

As illustration, let us do a simple estimation. Let us suppose that 1% of New York's population (200 thousand people) installs a KLJN cipher based network card in their PC and they are connected by a pair of wires in a chain-like fashion, see Figure 7. They are also connected to the internet and a central CS server with their regular Eternet card. Let us suppose that the distance between nearest neighbors is less then 1 km. Then it is very convenient to run the KLJN ciphers with a bandwidth 1000 bit/second or greater. Thus the theoretical speed of secret key generation and distribution to each computer in the network would be 200 Megabit/second. *Such a network and key generation/distribution speed is far beyond of not only the reach of any quantum communicator system but also that of the RSA and other software key distribution methods*. Of course, this speed is much higher than any PC can handle, thus the clock frequency of the KLJN ciphers could be reduced by orders of magnitude and the key generation and distribution could be still fast enough. The very important implication of this fact is that the requirements about the wire-pair connection of the KLJN cipher chain could be very low.

Alternatively, a much smaller network population could generate and distribute secret keys at sufficient speed. For example, 100 Units with 1000/bit/second clock frequency would generate and share a secret key by 100 kbit/second speed.

Note, the above-described speed holds for the current/voltage based defense only. More economical solutions will result in a speed reduction by a factor of 3 (see Sections 7,8).

**7. Enhancements for total security at economic design to avoid the "Mingesz" attacks**

One of us (Robert Mingesz) [11] has pointed out two important vulnerabilities of the network described above. In this section, we describe the problem and show one type of the possible solutions.

*i)* If during the initialization described in Section 5 at point *i)* a brief man-in-the-middle-attack takes place and *if the system is defended against the man-in-the-middle attack by the economic solution* described in section 3, the eavesdropper (with a small probability) may get the value of the secure bit because 4 bit eavesdropping is needed (on the average) to discover the man-in-the-middle.

*ii)* A much more serious problem is that when the eavesdropper attacks the regular network at Unit (*k*), and because that network is not totally secure (at least at the beginning) she may get the $G_{j,k}$ functions for each *j* location. But that set of $G_{j,k}$ functions is either the key or its inverse, depending on the local bit at Unit (*k*).

These problems must be fixed for the total security and this needs a more elaborate initialization resulting in a factor of 2 reduction of the network speed.

*iii)* Either the SC communicator port initialization/distribution must be done after each N-bit key generation or the man-in-the-middle defense should be the more expensive current/voltage based defense described in [3]. The fastest way of part of SC initialization/distribution is described here. During the secure port (re)distribution the Units should be run so that all ports change randomly and independently at the beginning of each clock period until a new port distribution is reached in accordance with the truth table, Table 3 (confer Section 5, Table 2). This method results in a new port distribution after 4/3 clock periods, on the average.

| **Situation of secure bits** | Low at $L_1^{M+1}$<br>High at $R_1^M$<br>and/or<br>Low at $L_2^{M+1}$<br>High at $R_2^M$ | Low at $R_1^M$<br>High at $L_1^{M+1}$<br>and/or<br>Low at $R_2^M$<br>High at $L_2^{M+1}$ | All other cases |
|---|---|---|---|
| **Resulting port control** | $R_1^M = R_2^M$<br>$L_1^{M+1} = -L_2^{M+1}$ | $R_1^M = -R_2^M$<br>$L_1^{M+1} = L_2^{M+1}$ | Do it again (No port distribution) |

**Table 3.** Truth table of the high-speed secure sharing of the port controls between the two nearest neighbors, Unit (M) and Unit (M+1), in the chain-like KJLN network. Only secure bit exchanges generate a new port distribution. The mean time needed is 4/3 clock periods because, from the 16 possibilities, 12 results in a new port distribution.

*iv)* As part of the initialization process, the very first N-bit secret key distribution has to be done in a slower way, without telecloning. Moreover, the Coordinator-server should also possess one Unit of the network. Let us suppose that the right ports of Unit (1) and the left ports of Unit (2) act as the first network-key generator. Unit (1)'s *R* ports and Unit (2)'s *L* ports generate, share and store an N-bit long key during N clock periods. At the same time, all other connected *L* and *R* ports also generate N-bit long keys for the communication between the nearest neighbors. Then by using the regular network and the key shared between Units (1) and (2), Unit (2) sends the N-bit long network-key to Unit (3). Then Unit (3) forwards the network-key to Unit (4) by using the regular network and the key generated between them, and so on... In this way, after N clock periods and then forwarding the keys in N-1 subsequent steps via the fast regular network, all Units, including the Coordinator-server, will learn the first, N-bit long network-key generated by Units (1) and (2). From that moment on, this key and its successors should be used for the communication between the Coordinator-server and the Units. The result is very close to a one-time-pad type encryption. Taking the New York example mentioned above, and assuming that one step of sending the network-key in the regular network to the nearest neighbor takes the same time as the clock period of the KLJN ciphers, this totally secure initialization would take 400 seconds. After this $\approx 7$ minutes initialization process, the network begins the fast and continuous production of totally secure keys. Due to the redistribution of ports after each N-bit key distribution, the speed of secure key generation and whole-distribution would be about the third of the speed mentioned above, about 70 Mbit/second.

## 8. The simplest and most economical network solution

As it is pointed out in 7/*iii*, the inexpensive defense solution with the required SC port redistribution will need and extra 4/3 clock period time at each network key generation. This requirement will slow down the key generation/distribution by more than a factor of 2. Practically, the network key generation will need 3 clock steps. Therefore, it is worthwhile to consider the use a much simpler network based on simple KLJN loops instead of the SC type complementary loops. Indeed, if on Figure 7, we use only single right and left ports the connected $R_k$ and $L_{k+1}$ ports represent a regular KLJN cipher, as described in [1]. They can switch randomly at each clock period and, on the average, every second clock period produces a secure bit. The network, its initialization and running are basically the same (with minor differences) as described for the SC cipher based network. This very simple system is at least as fast as the SC cipher based system with the economical defense (Section 3). Practically, every 3rd clock step will produce and distribute a network key of N bits. Therefore, the SC cipher based solution seems to have only one advantage: with the current/voltage based defense solution, it has the highest speed, about 3 times faster than that of the other solutions.

## 9. Remarks questions, perspectives

It is important to note that the network described in Figure 7 is very different from the basically *point-to-point* key distribution methods used by quantum communication and software solutions. Even though, the possibility of telecloning of quantum states to multiple receivers has been pointed out by van Loock and Braunstein [8] the fidelity is poor (<71%) and to reach an acceptable fidelity (>50%) the number of Units has to be limited to 30. Moreover, each Unit has to be connected to all the other Units by a separate communicator and separate lines, which means $N(N-1)$ communicator devices and $(N^2-N)/2$ lines indicating that the method is essentially a point-to-point communication type. For the New York example mentioned above, such a quantum telecloning solution requires about 200 *thousand communicator devices at each Unit* and that requires almost 40 billion communicator devices and nearly 40 billion optical cables. Moreover, these 40 billion connections are both short-distance and *long-distance* ones because we have to directly connect the farthest Units, too.

On the other hand, the KLJN-based network (Figure 7) requires only two communicator devices for each intermediate Units and one for the end Units. The intermediate Units have to be connected only with the two nearest neighbors and the end Units of only one neighbor. That requires only $2*(N-1)$ communicator devices and $(N-1)$ wires. In the New York example mentioned above, the two communicator devices at each Unit and two wires at each Unit makes only about 200 thousand communicator devices and about 200 thousand cable connections, moreover, these *connections are all of short distance type*, connecting only the nearest neighbors. If the regular network and the CS are fast enough compared to the KLJN clock, the whole network receives an N-bit long key at every second KLJN clock period. On the other hand, the whole network will receive the same key. Therefore, the system is totally secure only against external attacks: *hackers from outside* the network. Within the network, the security, which can be added to the network-key security, is only a regular network security protecting the information sent to/from to the CS Unit. Thus, we have the same level of security *against hackers within the network*, as regular networks do.

This special situation generates a lot of different open questions. For example:

*a)* What is the proper approach to encryption when we have this continuous high-speed key generation and simultaneous whole-network-key distribution?

*b)* Can the generated secure bits be used to increase the security of the internal network?

*c)* Higher dimensional network topologies and redundancy to protect against broken lines or Units down. This can, for example, be realized with server units with more than two ($L$ and $R$) ports. The ports $P_i(k)$ ($i = 1...Q$) of the $k$-th server can be connected to up to $Q$ different Units. The Coordinator-server would collect the logic relations and evaluate Eq. 1, accordingly.

d) Network redundancy and coding at higher dimensional topologies (?)

e) This network can alternatively be used to announce information to one or more Unit(s) or to the whole network simultaneously, in a totally secure way. Is there a need for this kind of solution?

## 10. Possible application of the idea in quantum communication

There are quantum key distribution schemes which produce entangled secure bits at the two ends of the line. Concerning the network solution described in Section 8, the same sequence of ideas, the same network topologies and initialization processes can be applied for such networks where the KLJN-based Units are replaced by quantum entangled state based Units. Such a network would produce the *multiple telecloning of classical bits via a quantum communicator networks without telecloning the quantum states*. This can (theoretically) be a 100% fidelity process and could be a large chain like network, for example. From these considerations, it is obvious that the telecloning of quantum states is an unnecessary luxury because it has many disadvantages, such as the need of separately connecting each units, and the very low fidelity. What is needed for practical applications: the secure transfer of the information bit. And that can be done also via the chain-like (or other) type of network of quantum channels and a regular network with the Coordinator-server. Because there are situations where wire connection is impossible but optical channel can be realized (like space communication with laser beam) this type of *quantum-based secure classical communication* may be find applications.

## 11. Conclusion

We have used electrically isolated KLJN ciphers to build enhanced ciphers and network. We introduced the Secure Complementary (SC) KLJN chiper for the top speed of secure key generation and exchange. We built network Units from two SC ciphers with random internal logic relations. We built a one-dimensional network using the network Units and a regular network with a Coordinator-server (CS). With such a network a high-speed, whole-network, one-step, secret key distribution can be achieved instead of the slow, point-to-point methods used by quantum communication or RSA-like key exchange methods. This is a teleportation-type multiple telecloning of the classical information bit because the information transfer can take place without the actual presence of the information bit at the intermediate points of the network. We have shown other, more economical and still fast-enough solutions, too.

In Table 4, a quick comparison between quantum communicators and the KLJN crypto system is shown.

## 12. Acknowledgments

The quantum part of Table 4 is a personal opinion of the authors based on assessments of information accessed/acquired. The authors urge the Reader to make his/her own assessments based on the scientific content of the present paper and papers on quantum communication in the literature (see [8] and similar papers).

| | **Quantum Comm.** | **KLJN Comm.** |
|---|---|---|
| **Physics behind the security** | Quantum (Fragile information bit) | Classical statistical (Robust information bit) |
| **Max. number of eavesdropped bits before 99% probability of eavesdropper detection** | Few thousand | 0-4 |
| **Vulnerability against the man-in-the-middle attack** | Usually yes | No |
| **Information leak below the eavesdropper detection radar (eavesdropper hiding in noise)** | >1% | 0.01% or less is easily reachable |
| **Ultimate speed-cut-off versus range** | Exponential cut-off | 1/range cut-off |
| **Network key distribution** | No. Only point-to-point | Yes. Whole-network key distribution within two clock periods |
| **Telecloning** | Yes, with fidelity < 71% | Yes, with 100% fidelity |
| **Network telecloning in one step. Number of units $N$.** | Only if all Units are *directly connected to each other* which needs $(N^2-N)/2$ connections. Acceptable fidelity (>50%) only for $N \approx 30$, or less. | Yes, even if the Units connected as an arbitrarily large ($N \to \infty$) *chain network* (needs only $N-1$ connections), within two clock periods. |
| **Vibration resistant** | No | Yes |
| **Shock resistance** | Poor | Excellent |
| **Dust resistant** | No | Yes |
| **Microelectronic integrated parallel multi-line (>100) driver chip** | No | Yes |
| **Low-power consumption** | No | Yes |
| **Communicator as a computer card** | No | Yes |
| **Price** | High | Low |

**Table 4.** Quick comparison of quantum communicators (where the quantum states are transferred) with the KLJN system (March 11, 2006). See disclaimer about the quantum part in the Acknowledgements.